\documentclass[12pt,onecolumn]{IEEEtran} % arxiv Mode
\usepackage{subfigure}
\usepackage{graphicx}
\usepackage{color}
\usepackage{times}
\usepackage{booktabs}
\usepackage{multirow}
\usepackage{rotating}
\usepackage{multicol}
\usepackage{dsfont}
\usepackage{amssymb}
\usepackage{float}
\usepackage{cite}
\usepackage{eurosym}
\usepackage{amsmath}
\usepackage{setspace}
\usepackage{hhline}
\usepackage{psfrag}
\usepackage{graphicx}
\usepackage{subfigure}
\usepackage{acronym}
\usepackage{flushend}
\usepackage{booktabs}
\acrodef{hpfc}[HPFC]{Hourly Price Forward Curve}
\acrodef{pfc}[PFC]{Price Forward Curve}
\acrodef{hdd}[HDD]{heating degree day}
\acrodef{cdd}[CDD]{cooling degree day}
\acrodef{vpp}[VPP]{virtual power plants}
\acrodef{epex}[EPEX]{European Power Exchange}
\acrodef{eex}[EEX]{European Energy Exchange}
\acrodef{omel}[OMEL]{Operadora del Mercado Espa\~nol de Electricidad}
\acrodef{res}[RES]{renewable energy sources}
\acrodef{fit}[FIT]{feed-in tariff}
\acrodef{eeg}[EEG]{"Erneuerbare-Energien-Gesetz"}
\acrodef{seg}[SEG]{"Stromeinspeisungsgesetz"}
\acrodef{pv}[PV]{photovoltaic}
\acrodef{tso}[TSO]{transmission system operator}
\acrodef{otc}[OTC]{over-the-counter}
\acrodef{pshp}[PSHP]{pumped storage hydro plant}
\acrodef{entsoe}[ENTSO-E]{European Network of Transmission System Operators for Electricity}

\acrodef{ode}[ODE]{ordinary differential equation}
\acrodef{lp}[LP]{linear program}
\acrodef{qp}[QP]{quadratic program}
\acrodef{fft}[FFT]{Fast Fourier Transformation}

\acrodef{mape}[MAPE]{mean absolute prediction error}
\acrodef{mae}[MAE]{mean absolute error}
\acrodef{mse}[MSE]{mean square error}
\acrodef{sma}[SMA]{simple moving average}
\acrodef{ewma}[EWMA]{exponentially weighted moving average}
\acrodef{ols}[OLS]{ordinary least squares}
\acrodef{lad}[LAD]{least absolute deviation}
\acrodef{ladlasso}[LAD-LASSO]{Least Absolute Deviation with Least Absolute Shrinkage and Selection Operator}
\acrodef{lasso}[LASSO]{Least Absolute Shrinkage and Selection Operator}
\acrodef{sde}[SDE]{stochastic differential equation}
\acrodef{mle}[MLE]{Maximum Likelihood Estimator}
\acrodef{aic}[AIC]{Akaike Information Criterion}
\acrodef{bic}[BIC]{Baysian Information Criterion}
\acrodef{ar}[AR]{autoregressive}
\acrodef{arx}[ARX]{autoregressive with external inputs}
\acrodef{arma}[ARMA]{autoregressive moving average}
\acrodef{arima}[ARIMA]{autoregressive integrated moving average}
\acrodef{wn}[WN]{white noise}
\acrodef{moc}[MOC]{Merit-Order Curve}
\acrodef{rmse}[RMSE]{root mean square error}

\acrodef{fl}[FL]{fuzzy logic}
\acrodef{svm}[SVM]{Support Vector Machine}
\acrodef{lssvm}[LSSVM]{Least Squares Support Vector Machine}
\acrodef{elm}[ELM]{Extreme Learning Machine}

\acrodef{var}[VaR]{Value-at-Risk}
\acrodef{cvar}[CVaR]{conditional Value-at-Risk}
\acrodef{par}[PaR]{Profit-at-Risk}
\acrodef{cpar}[CPaR]{conditional Profit-at-Risk}
\acrodef{nn}[NN]{neuronal networks}
\acrodef{ann}[ANN]{artificial neural networks}

\acrodef{gdp}[GDP]{gross domestic product}
\acrodef{oecd}[OECD]{Economic Co-operation and Development}

\acrodef{kti}[KTI]{Swiss Innovation Promotion Agency}
\acrodef{sqg}[SQG]{\textrm{swissQuant} Group AG}

\begin{document}
\title{\textbf{Grid Integration Costs of\\Fluctuating Renewable Energy Sources}}
%\title{Grid-integration Costs of Fluctuating Renewable Energy Sources}
%\author{\IEEEauthorblockN{Jonas M\"uller}
%\IEEEauthorblockA{Power Systems Laboratory, ETH Z\"urich, Switzerland\\
%Email: muellejo@student.ethz.ch}\vspace{8pt}
%\IEEEauthorblockN{Andreas Ulbig}
%\IEEEauthorblockA{Power Systems Laboratory, ETH Z\"urich, Switzerland\\
%Email: ulbig@eeh.ee.ethz.ch}\vspace{8pt}
%\and
%\IEEEauthorblockN{Marcus Hildmann}
%\IEEEauthorblockA{Power Systems Laboratory, ETH Z\"urich, Switzerland\\
%Email: hildmann@eeh.ee.ethz.ch}\vspace{8pt}
%\IEEEauthorblockN{G\"oran Andersson}
%\IEEEauthorblockA{Power Systems Laboratory, ETH Z\"urich, Switzerland\\
%Email: andersson@eeh.ee.ethz.ch}
%%\and
%}
\author{
	Jonas M\"uller, Marcus Hildmann, Andreas Ulbig and G\"oran Andersson\\
	%\IEEEauthorblockN{Andreas Ulbig$^*$ and G\"{o}ran Andersson}
	Power Systems Laboratory, ETH Zurich\\
	Zurich, Switzerland\\
	{\textmd muellejo@ethz.ch \quad \textmd hildmann$\;$\textbar$\;$ulbig$\;$\textbar$\;$andersson$\,$@$\,$eeh.ee.ethz.ch}
\thanks{\footnotesize J.~M\"uller is currently a master student at ETH Zurich. His related master thesis was conducted at the ETH Power Systems Laboratory, supervised by the co-authors.}
}
\maketitle
% \begin{center} \bfseries EDICS Category: 3-BBN    §D \end{center}
% \fi
%\IEEEpeerreviewmaketitle

% \doublespacing

\begin{abstract}
The grid integration of intermittent \ac{res} causes costs for grid operators due to forecast uncertainty and the resulting production schedule mismatches. These so-called profile service costs are marginal cost components and can be understood as an insurance fee against \ac{res} production schedule uncertainty that the system operator incurs due to the obligation to always provide sufficient control reserve capacity for power imbalance mitigation.
This paper studies the situation for the German power system and the existing German \ac{res} support schemes. The profile service costs incurred by German \acp{tso} are quantified and means for cost reduction are discussed. In general, profile service costs are dependent on the \ac{res} prediction error and the specific workings of the power markets via which the prediction error is balanced. This paper shows both how the prediction error can be reduced in daily operation as well as how profile service costs can be reduced via optimization against power markets and/or active curtailment of \ac{res} generation.
\end{abstract}

\vspace{0.25cm}
%\begin{IEEEkeywords}
%\end{IEEEkeywords} 
% reset acronym counter from abstract
\acresetall
\section{Introduction}
\label{sec:introduction}

Since the early 1980s, government support schemes with the specific goal of promoting large-scale deployment of \ac{res} were introduced in many countries worldwide. The German Renewable Energy Act, \ac{eeg}, a well-known support scheme, provides a favorable \ac{fit} for a variety of \ac{res} since the year~2000 and builds on the good experience with its predecessor, the \emph{Stromeinspeisungsgesetz} from 1991. It gives priority to electric power in-feed from \ac{res} over power in-feed from conventional power plants,~i.e., fossil-fueled and nuclear and large, hydro-based power plants. This favorable investment environment has led to a massive build-up notably of wind \& \ac{pv} generation in Germany~\cite{AGEB2013, BDEW2013}. By year-end~2013, the installed wind and \ac{pv} power capacities were around 33 GW and 39 GW respectively~\cite{GmbH2013}. The original goal of the \ac{fit},~i.e. inciting large-scale \ac{res} deployment has thus been achieved.

With a combined installed power capacity of \ac{res} units of more than 70~GW, somewhat more than the average load demand in the German power system, of ~63-68~GW, and significant annual RES energy shares, about 15\% combined, wind \& \ac{pv} units can no longer be treated as exotic, marginal electricity sources.
The current \ac{res} production already has significant effects on the power market, notably in the form of the so-called \emph{merit-order effect}. Especially the decoupling of spot market prices and \ac{res} in-feed due to \ac{fit} regulations, results in lower average spot price levels and also in negative spot prices for several hours each month. One effect of this is that flexible power plants such as gas-fired units cannot be operated profitably because peak spot prices are too often below their marginal operation costs. Another effect is that due to the also associated spread between peak and base base day-ahead prices, energy storage facilities, primarily \acp{pshp} cannot be operated profitably either~\cite{Hildmann2011a}.
%On the other hand, the provision of independent payments from \ac{fit} for 20 years results in high costs for the electricity consumers. %The so called EEG Umlage is 0.036 \euro/kWh in 2012 and will rise to 0.053 \euro/kWh in 2013, which reflects roughly 20\% of the end consumer tariff.
As a consequence, the production structure is about to be transformed from mainly centralized large power plants to a decentralized structure of small \ac{res} plants. This has effects on the operation of the grid and of the electricity markets including significant costs for \ac{res} grid integration; in addition to the in-feed tariffs.
% , as shown in Fig.~\ref{fig:installedCapacity}.
%\begin{figure}[bh]
%	\centering
%	\vspace{-0.3cm}
%     %\centerline{\psfig{figure=ComparisonPeakShapeYearly.eps,width=\linewidth} }
%    \psfrag{x1}[cc][cc]{\scriptsize\shortstack{Year}}
%    \psfrag{y1}[cc][cc]{\scriptsize\shortstack{GW}}
%    \includegraphics[width=0.475\textwidth]{installedCapacity.eps}
%    %\centerline{\psfig{figure=matlab/modifiedAskCurve, width=1\linewidth} }
%    \vspace{0.10cm}
%    \caption{Evolution of installed wind \& \ac{pv} capacity in Germany.}
%    \label{fig:installedCapacity}
%    \vspace{-0.25cm}
%\end{figure}
%\begin{figure}[h]
%	\centering
%	\vspace{-0.3cm}
%    %\centerline{\psfig{figure=ComparisonPeakShapeYearly.eps,width=\linewidth} }
%    \psfrag{x1}[cc][cc]{\scriptsize\shortstack{Year}}
%    \psfrag{y1}[cc][cc]{\scriptsize\shortstack{TWh}}
%    \includegraphics[width=0.425\textwidth]{overallProduction2012.eps}
%    %\centerline{\psfig{figure=matlab/modifiedAskCurve, width=1\linewidth} }
% 	\vspace{-0.20cm}
%    \caption{Evolution of annual in-feed of wind and \ac{pv} units in Germany.}
%    \label{fig:overallProduction2012}
%    \vspace{-0.3cm}
%\end{figure}

The remainder of this paper is outlined as follows: Section~\ref{sec:cost_structure} gives an overview of the relevant power market structures and cost drivers for \ac{res} grid integration. Section~\ref{sec:marketData} briefly discusses the employed data, which is further on analyzed in~Section~\ref{sec:analysis}. Finally, Section~\ref{sec:results} presents the key results.

\section{Cost Structure of Renewable Energy Sources}
\label{sec:cost_structure}

\subsection{Power Market Structures and Cost Drivers}
\begin{figure}[!htb]
\centering
\includegraphics[trim = 0cm  0.25cm  0cm  0.25cm, clip=true, width=0.65\textwidth]{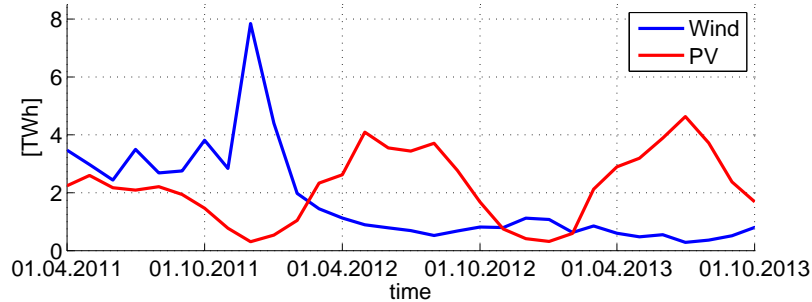}
\caption{Wind \& \ac{pv} volumes sold via the \ac{fit} scheme \cite{GmbH2013}.}
\label{fig:fitChanges}
\vspace{-0.25cm}
\end{figure}

\begin{figure}[!htb]
\centering
\includegraphics[trim = 0cm  0.25cm  0cm  0.25cm, clip=true, width=0.65\textwidth]{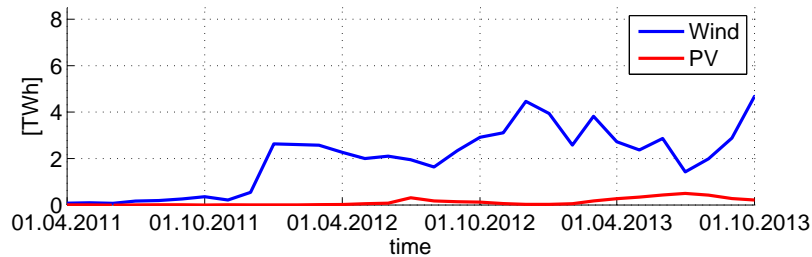}
\caption{Wind \& \ac{pv} volumes sold via direct marketing \cite{GmbH2013}.}
\label{fig:directMarketingChanges}
\vspace{-0.25cm}
\end{figure}

In Germany, the power market is organized as an area pricing market. Thus, the price model operates on the assumption of zero congestion. This results in a full decoupling of power price and production region, which brings the need for re-dispatch measures to compensate for transmission grid congestions.\\
The support scheme for most of \ac{res} in Germany is two-fold (since Jan. 2012): A producer can choose between \ac{fit} and a premium for direct sales. The constant \ac{fit} for \ac{res} production pays the producer a fixed amount of money. Furthermore the \ac{fit} regulations guarantees an in-feed priority over conventional generation \cite{HildmannPESGM12HPFC}. The power generated under \ac{fit} support is sold via the \ac{tso}, which also covers the \ac{res} forecast error. On the other hand, the direct sale is organized via third party \cite{Vorschriften2012}. It does not include guaranteed sell of the power and the forecast error must be handled by the third party. The \ac{eeg} states that the producers can freely switch between support schemes every month. Fig \ref{fig:fitChanges} and \ref{fig:directMarketingChanges} show the participation over time in the two support schemes. In 2012 the direct sales support scheme became preferable for wind plant operators. This is a result of the better forecast quality for wind production based on the strong auto-correlation of the wind in-feed.\\
The support schemes for \ac{res} and the decoupling of the electricity market from the underlying grid structures both result in costs for the energy consumer.

\subsubsection{Costs at Energy Only Markets}\label{sec:costDayAhead}
\begin{figure}[!htb]
\centering
\includegraphics[width=0.65\textwidth]{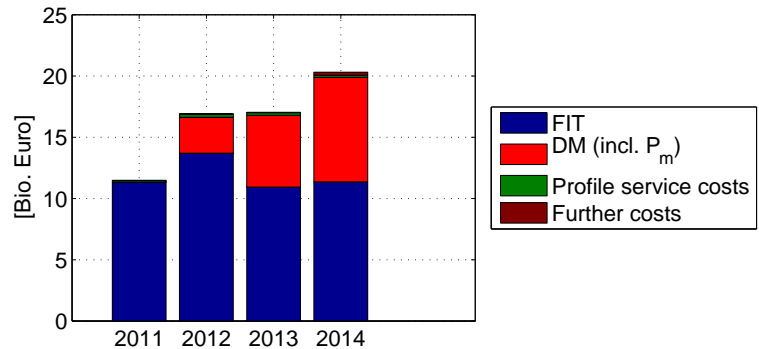}
\caption{EEG difference costs 2010-2013 and the estimation for 2014 \cite{EEGKWK2014}.}
\label{fig:eexCostStruct}
\vspace{-0.2cm}
\end{figure}

In Germany, two energy only markets exist, the day-ahead market and the intra-day market. Both markets are affected by the deployment of \ac{res} and serve different purposes. While the day-ahead market is the main procurement market, the intra-day market is used to settle the majority of the day-ahead forecast error before the use of ancillary services. For \ac{fit} supported \ac{res}, the  direct cost at the day-ahead market is the gap between the guaranteed \ac{fit} and the achieved market prices plus the respective variable profile service costs for the \ac{fit} production. For the direct sales the cost are the market premium including a management premium for the direct marketing participants. These direct costs are directly transferred to the electricity customer via a \ac{res} allocation charge the so called "EEG-Umlage". As a result of excess supply, priority of \ac{res} in-feed and the lack of competitive market clearing for direct marketing the so-called merit-order effect lowered the market prices significantly in recent times. To ensure the settlement of \ac{res} in the auction, the price-independent \ac{fit} supported power enters the market with the allowed minimum bid of $-3000\frac{\textrm{\euro}}{\textrm{MWh}}$. The direct sales \ac{res} can technically enter the market with minus the market premium plus a small balancing error premium. Both effects shift the merit-order-curve to the right and lower the achieved day-ahead market price. The merit-order-effect is discussed in detail in \cite{ResWorkingPaper2014}. The total paid "EEG-Umlage" in Germany alone was almost 17 billion euros in 2012 and 2013 and estimated almost 20 billion in 2014 as shown in Fig. \ref{fig:eexCostStruct}.\\
The intra-day market provide continues trading until 45 minutes before delivery. Compared to the pay-as-settled day-ahead market, the intra-day market is a pay-as-bid market. Since the market closing is only 45 minutes before delivery, the majority of the prediction errors can be settled on the intra-day market. Compared to the day-ahead market, the costs for balancing actions at the intra-day market are not part of the \ac{res} allocation charge but a component in the grid tariff.

\subsubsection{Costs on Ancillary Services Markets, Structural Changes \& Effects}
%As discussed in Section \ref{sec:costDayAhead} the support scheme results in direct costs from the the day-ahead market settlement. 
%Compared to the compensation for \ac{fit}, the cost increases for ancillary services are less transparent.
The transformation from centralized production to a decentralized production by small units has considerable ramification, on the power grid. The location mismatch of production and demand, the uncertain character of wind and \ac{pv} and the shutdown off old, non-profitable, powerplants make ancillary services and re-dispatch more complex and therefore more expensive.

\begin{figure}[!htb]
\centering
\includegraphics[width=0.65\textwidth]{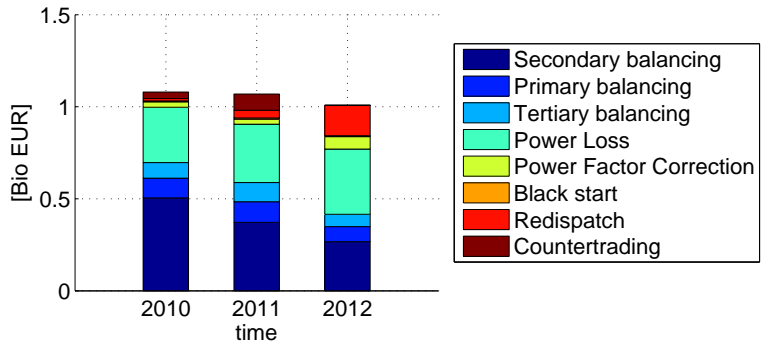}
\caption{Change cost for different uses of ancillary services \cite{Bundesnetzagentur2013}.}
\label{fig:furtherCostFig}
\vspace{-0.2cm}
\end{figure}

Fig. \ref{fig:furtherCostFig} shows that the costs of ancillary services over the years 2010, 2011 and 2012 were almost constant at 1 billion euros per year \cite{Bundesnetzagentur2013}. Anyhow, the composition of ancillary services changed over the years. The cost for secondary control (red) has decreased significantly as a result of the settlement of the forecast error (load and \ac{res} forecast) at the intra-day market. On the other hand, the major cost increase over the last three years is the compensation of power loss (cyan) in the grid and the re-dispatch measures. In the past, the majority of production was fed into the transmission grid, with low power loss, the areas of large demand close to the production location and proper grid design. With decentralized production the power flow goes from the low voltage levels via the transmission grid, to other low voltage distribution grid areas. More over, large \ac{res} production sources like offshore wind parks, are typically far away from the major location of power consumption. Power loss compensation and re-dispatch are performed by the \acp{tso} and are billed to the customers via grid tariffs. While most of the standard prediction errors, such as load prediction errors stay approximately constant, the wind and \ac{pv} generated in-feed has increased significantly over time \cite{ResWorkingPaper2014}.\\

%% END OF FILE 
\section{Data Section}\label{sec:marketData}
As a result of several transparency acts, the data for power markets, energy-only as well as ancillary services, has in recent years become available in good quality. As required by the \ac{eeg}, the prediction of wind and \ac{pv} in-feed, as well as the realized in-feed have to be published. Recently \ac{moc} data became available not only for the energy-only markets but also for the ancillary services markets.

\begin{table}[!htb]
\caption{Summary of Market Data}
\label{tab:availableDataEnergyMarket}
\centering
\begin{tabular}{lcc}
\hline \hline
% here begin the table head
Data Time-Series                  & Available since  & Source\\
\hline
% here begin the supply data
Wind power in-feed $W(t)$ & 2011/04 & \cite{EEGKWK2014}\\
Solar power in-feed $R(t)$ & 2011/04 & \cite{EEGKWK2014}\\
Wind in-feed pred $W(t)'$ & 2011/04 & \cite{EEGKWK2014}\\
Solar in-feed pred $R(t)'$ & 2011/04 & \cite{EEGKWK2014}\\
\acs{epex} Day-ahead prices $S(t)$ & $2002/01$ & \cite{EPEX}\\
\acs{epex} Intra-day prices $I(t)$ & 2006/10 & \cite{EPEX}\\
Balancing Energy Price $P_{\text{bal}}$ & 2001/02 & \cite{Amprion}\\
Secondary Control \acs{moc} & 2007/12 & \cite{Regelleistung} \\
Tertiary Control \acs{moc} & 2007/12 & \cite{Regelleistung} \\
Management premium $P_m$ & 01/2012 & \cite{Vorschriften2012}\\
Other on \acs{eeg} costs & 01/2011 & \cite{EEGKWK2014}\\
\hline \hline
\vspace{-0.35cm}
\end{tabular}
\end{table}
The used data, the denomination of the variable and the source are given in Table \ref{tab:availableDataEnergyMarket}. All data referring to load has 15 minute resolution, while price data is provided with hourly resolution.
Since Germany is divided in four \ac{tso} areas, all data is \ac{tso} specific. If the data is \ac{tso} specific, the symbol carries the subscript $_{\textrm{\ac{tso}}}$.\\
In general, the market design, availability and transparency and \ac{res} support schemes differ from country to country. By all means Germany is a good study case for a market with a high \ac{res} in-feed with good data availability.

\section{Analysis of Profile Service Costs}\label{sec:analysis}
In this section we derive and analyse the profile service costs based on the available data. The day ahead forecast (08:00 AM \cite{Lenzi2013}) of intermittent \ac{res} generates an error $\epsilon(t)$ given by
% Profile service costs are costs arising from the balancing of intermittent \ac{res}.
\begin{equation}
\epsilon(t) = W(t)' - W(t) + R(t)' - R(t),
\end{equation}
where $W(t)'$ and $R(t)'$ are the day ahead generation forecasts for wind and \ac{pv} and $W(t)$ and $R(t)$ the effective generation. The prediction error per \ac{tso}, $\epsilon_{\text{TSO}}$, is defined analog as
\begin{equation}
\epsilon_{\text{TSO}}(t) = W_{\text{TSO}}(t)' - W_{\text{TSO}}(t) + R_{\text{TSO}}(t)' - R_{\text{TSO}}(t).
\label{eq:epsilonTSO}
\end{equation}
Because demand and supply must always match, $\epsilon_{\text{TSO}}(t)$ has to be leveled by the \ac{tso}. The prediction error known 45 minutes ahead is settled on the intra-day market. The remaining difference will be covered by secondary or tertiary control reserve. Therefore $\epsilon_{\text{TSO}}(t)$ can also be written as
\begin{equation}
\epsilon_{\text{TSO}}(t) = V_{\text{intraday}}(t) + V_{\text{bal}}(t),
\label{eq:volumes}
\end{equation}
where $V_{\text{intraday}}(t)$ is the volume traded at the intra-day market and $V_{\text{bal}}t)$ is the volume balanced via secondary or tertiary control. The time index $t$ refers to the delivery period which, in the case of the intra-day market, corresponds with the delivery time on the ancillary services market.

\subsection{Cost Structure}
Because only the intra-day market and balancing reserve is allowed to be used to cover the forecast errors in the \ac{tso} \ac{res} balancing area, the profile service cost $C(t)$ for time $t$ can be stated as the cost for intra-day market and balancing energy cost as
\begin{equation}
C(t) = C_{\text{intraday}}(t) + C_{\text{bal}}(t).
\label{eq:costs}
\end{equation}

\subsubsection{Intra-Day Market Balancing Costs}
\begin{figure}[!htb]
\centering
\includegraphics[width=0.65\textwidth, draft=false]{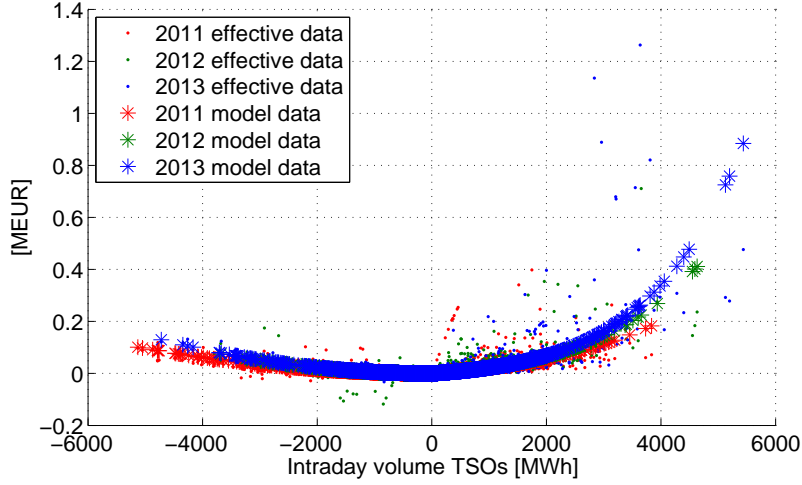}
\caption{Modelled hourly intra-day market costs for the German \acp{tso} in comparison to real data.}
\label{fig:Intraday costs}
\end{figure}

The intra-day market costs $C_{\text{intraday}}(t)$ are the difference between the revenue, or costs, on the intra-day market minus the foregone, or achieved, revenue at the \ac{epex} day-ahead market and is given as
\begin{equation}
C_{\text{intraday}}(t) = V_{\text{intraday}}(t)*(I(t)-S(t)),\\
\end{equation}
where $I(t)$ are the intra-day prices and $S(t)$ the day-ahead prices. The costs at the intra-day market $C_{\text{intraday}}(t)$ are positive for the buyer side if $V_{\text{intraday}}(t)$ is positive and $I(t)$ is bigger than $S(t)$. For the seller side the costs are positive, if $V_{\text{intraday}}(t)$ is negative and $I(t)$ is smaller than $S(t)$. Fig.~\ref{fig:Intraday costs} shows the dependency between $V_{\text{intraday}}(t)$ and $I(t)-S(t)$ of \ac{epex} market data. The positive cost show an exponential dependency between $I(t)-S(t)$ and positive intra-day volume $V_{\text{intraday}}(t)$ and a linear correlation between $I(t)-S(t)$ and negative intra-day $V_{\text{intraday}}(t)$.

Two conclusions can be drawn from the analysis and modelling of data. First, purchase on the \ac{epex} intra-day market is more expensive than sales of excessive energy as a result of the exponential function shape shown in Fig. \ref{fig:Intraday costs}. Second, modelling of purchase costs may underestimate the extreme costs arising in situations of market supply scarcity as indicated by the positive outliers shown in Fig. \ref{fig:Intraday costs}.

\subsubsection{Costs for Balancing energy}
Accordingly to the costs of the intra-day energy, the cost for balancing energy $C_{bal}(t)$ is defined as the product of balanced volume and difference of the price for balancing energy $P_{\text{bal}}(t)$ and the \ac{epex} day-ahead market $S(t)$. This gives
\begin{equation}
C_{\text{bal}}(t) = V_{\text{bal}}(t)*(P_{\text{bal}}(t)-S(t)).
\end{equation}
The price of balancing energy $P_{\text{bal}}(t)$ is defined as the costs of activation of balancing reserve $C_{\text{saldo}}(t)$ over the balancing saldo $RZ_{\text{saldo}}(t)$ for all balancing regions in Germany \cite{AEPmodell} as
\begin{equation}
P_{\text{bal}}(t) = \frac{C_{\text{saldo}}(t)}{RZ_{\text{saldo}}(t)}.
\label{eq:AEP}
\end{equation}

To avoid extreme prices for very low $RZ_{\text{saldo}}(t)$ and other price anomalies, $P_{\text{bal}}(t)$ is restricted in certain cases. For more information please refer to \cite{AEPmodell}.
The balancing energy saldo $RZ_{\text{saldo}}$ is defined as the sum of all activated secondary- and tertiary control reserves $V_{\text{sec},i}(t)$ and $V_{\text{ter},k}(t)$ as
\begin{equation}
RZ_{\text{saldo}}(t) = \sum_{i=1}^pV_{\text{sec},i}(t)+\sum_{k=1}^mV_{\text{ter},k}(t),
\end{equation}
where $p$ and $m$ are the index of the last considered bid of secondary and tertiary control reserve respectively.

The $RZ_{\text{saldo}}(t)$ consist of the following variables:
\begin{itemize}
\item $V_{\text{bal}}$ (residual prediction error of the \acp{tso})
\item residual prediction error from direct marketing
\item residual demand prediction error
\item demand fluctuations
\item unexpected power plant shutdowns
\end{itemize}

%Since the strong growth of \ac{res}, a linear dependency ($\rho$ = 0.51) can be drawn between $V_{\text{bal}}$ and $RZ_{\text{saldo}}$ averaged over the last three years. To account for uncertainty in the correlation, a Laplace distributed number has been added to the mean value correlation. According to data analysis, $\sum_{k=1}^mV_{\text{ter},k}$ can be set to zero until $RZ_{\text{saldo}}$ has breached $0.8*V_{sec,max}$.

\begin{figure}[!htb]
\centering
\includegraphics[width=0.65\textwidth, draft=false]{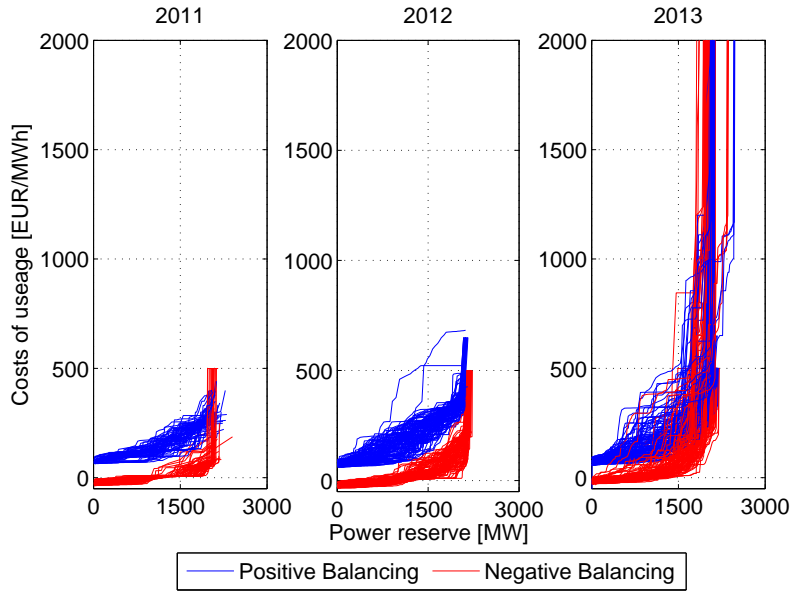}
\caption{Secondary control reserve merit-order of activation energy price \cite{Regelleistung}.}
\label{fig:MeritSec}
\vspace{-0.35cm}
\end{figure}

\begin{figure}[!htb]
\centering
\includegraphics[width=0.65\textwidth, draft=false]{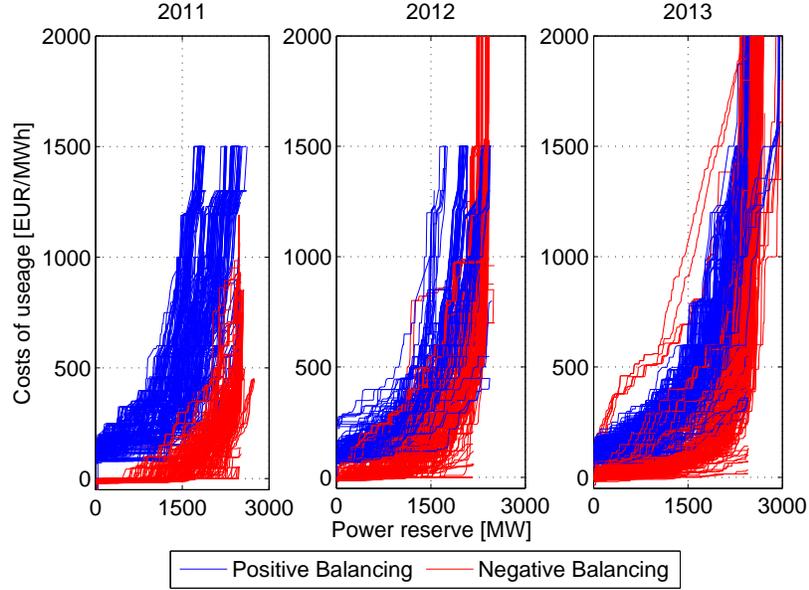}
\caption{Secondary control reserve merit-order of activation energy price \cite{Regelleistung}.}
\label{fig:MeritTer}
\vspace{-0.35cm}
\end{figure}

\begin{figure}[!htbp]
\centering
{\includegraphics[trim = 0.1cm 0cm 0.1cm 0cm, clip=true, angle=0, width=0.485\textwidth, keepaspectratio, draft=false]{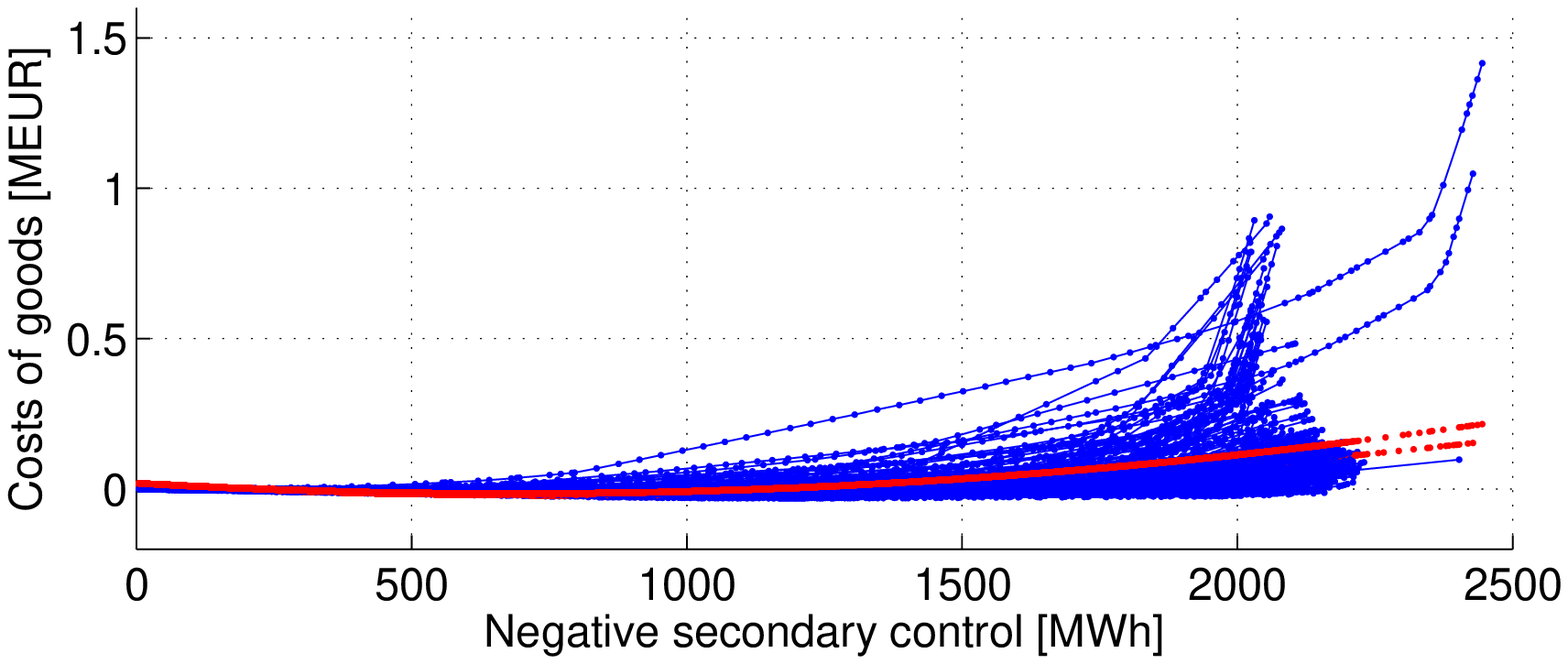}}
{\includegraphics[trim = 0.1cm 0cm 0.1cm 0cm, clip=true, angle=0, width=0.485\textwidth, keepaspectratio, draft=false]{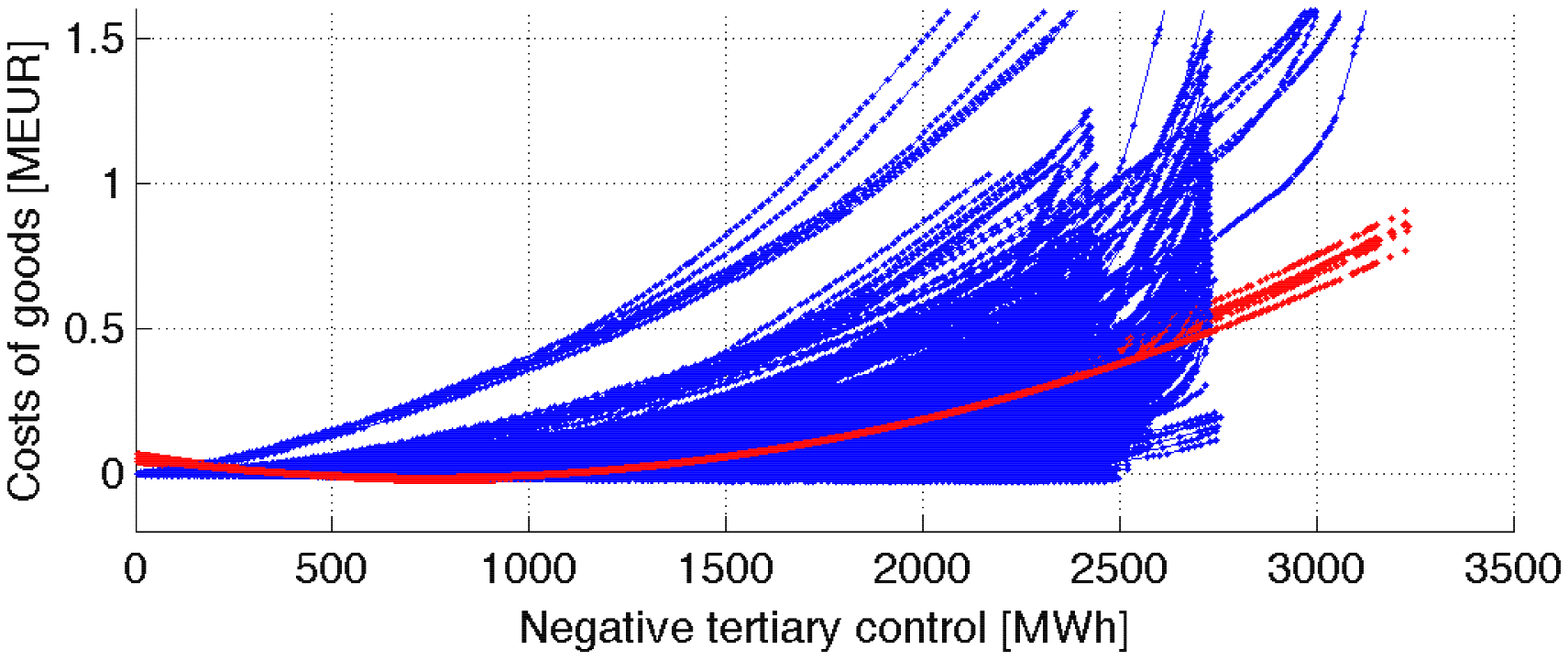}}
{\includegraphics[trim = 0.1cm 0cm 0.1cm 0cm, clip=true, angle=0, width=0.485\textwidth, keepaspectratio, draft=false]{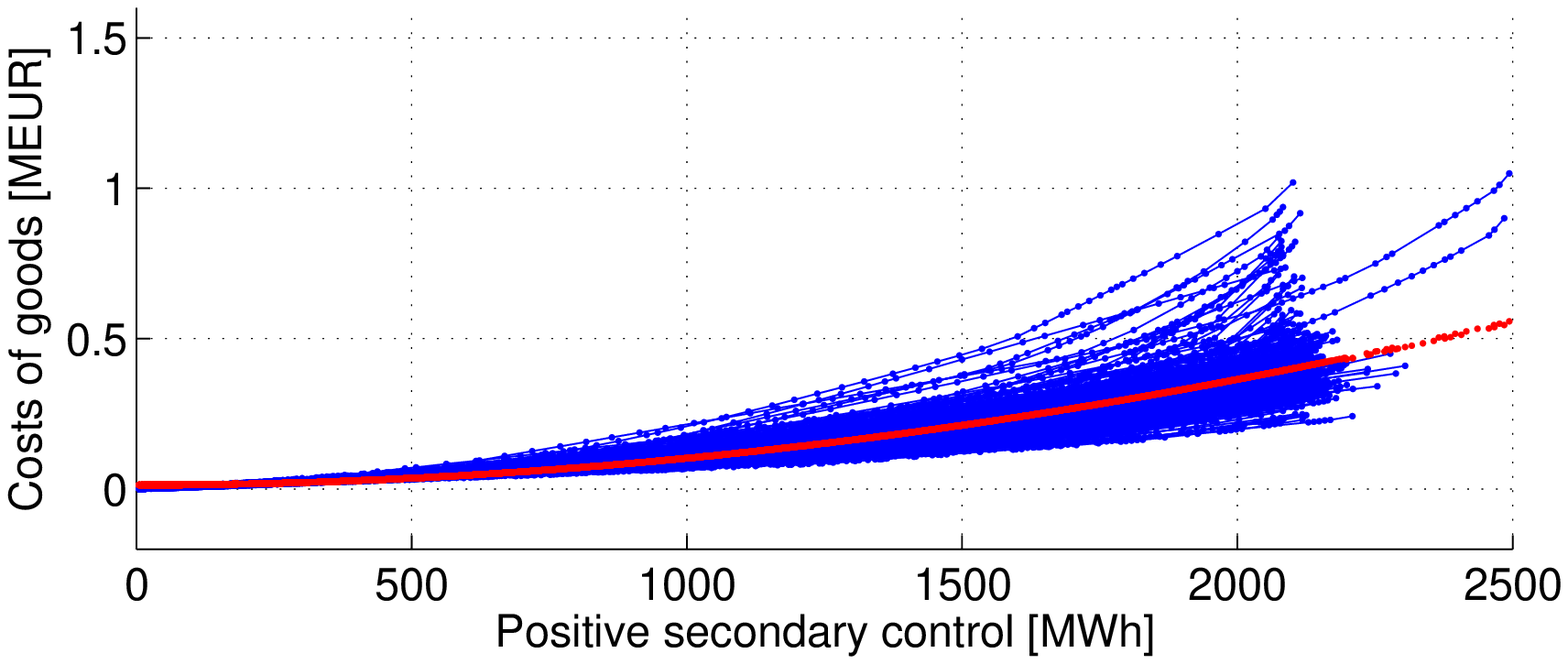}}
{\includegraphics[trim = 0.1cm 0cm 0.1cm 0cm, clip=true, angle=0, width=0.485\textwidth, keepaspectratio, draft=false]{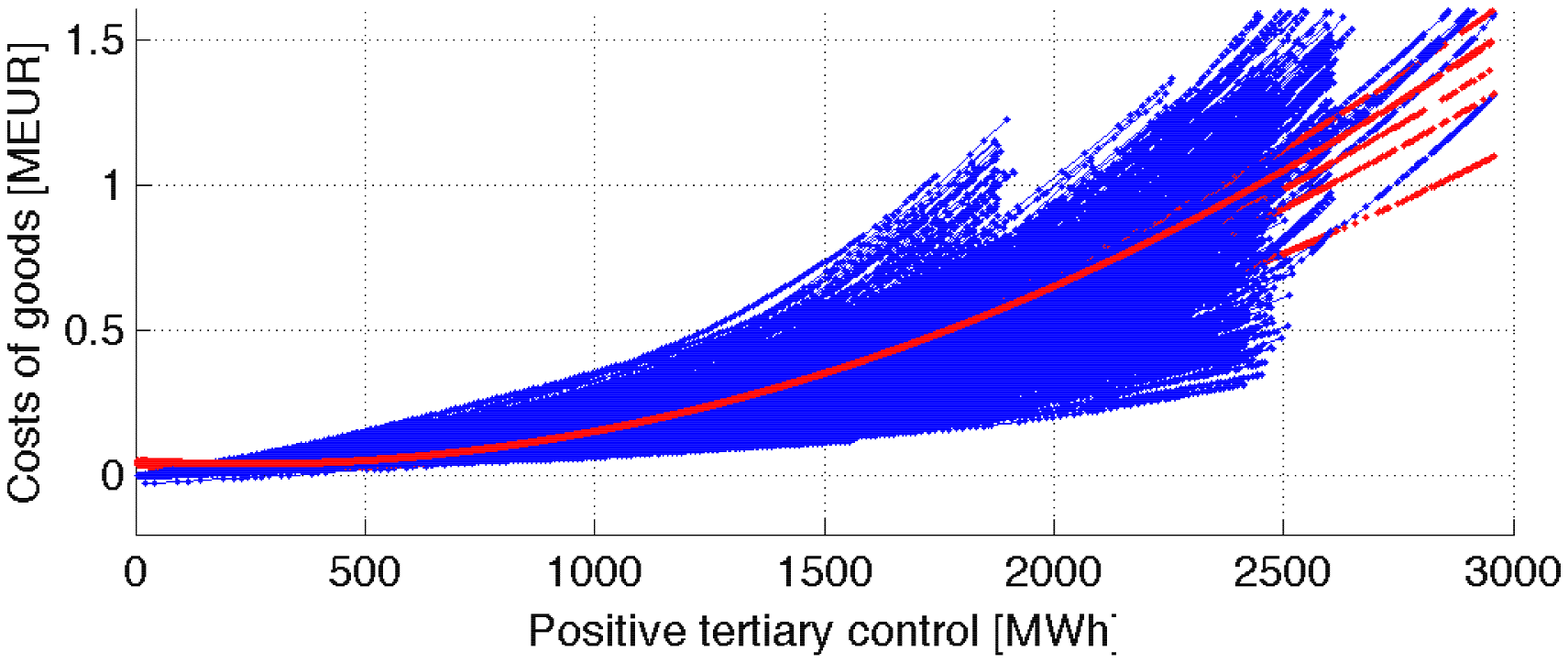}}
\caption{The four modelled costs of activation of secondary and tertiary control reserve.}
\label{fig:ModelBalact}
\end{figure}

The costs $C_{\text{saldo}}(t)$ can be defined as the sum of the costs of the activation of secondary and tertiary control as
\begin{equation}
C_{\text{saldo}}(t) = \sum_{i=1}^pV_{\text{sec},i}(t)*P_{\text{sec},i}(t)+\sum_{k=1}^mV_{\text{ter},k}(t)*P_{\text{ter},k}(t),
\end{equation}
where $i$ is the index of a balancing capacity which offers secondary control reserve $V_{\text{sec},i}(t)$ for an activation price $P_{\text{sec},i}(t)$, and $k$ is the index of a balancing capacity which offers tertiary control $V_{\text{ter},k}(t)$ for an activation price $P_{\text{ter},k}(t)$. Fig. \ref{fig:MeritSec} and Fig. \ref{fig:MeritTer} show the 15 minutes \ac{moc} for the years 2011 to 2013 for secondary and tertiary control \cite{Regelleistung}.
We train a model based on a quadratic function to model $P_{\text{bal}}$ as a function of $V_{\text{bal}}(t)$. The modelling is performed on aggregated data of all products corresponding to either of the four groups, negative secondary control reserve, positive secondary control reserve, negative tertiary control reserve and positive tertiary control reserve. Fig. \ref{fig:ModelBalact} shows the fit of the quadratic function (red) to the aggregated \acp{moc} data from the years 2011 to 2013.

\begin{figure}[!htb]
\centering
\includegraphics[width=0.65\textwidth, draft=false]{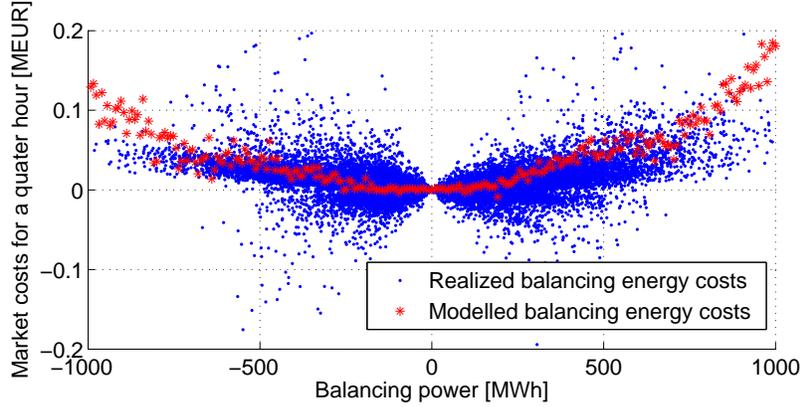}
\caption{Modelled balancing costs in comparison to effective balancing costs.}
\label{fig:Modelledbalancingcosts}
\end{figure}

The total balancing need $RZ_{\text{saldo}}(t)$ can be derived as a function of $V_{\text{bal}}(t)$. Using the four cost functions for the activation of control reserve to shown in Fig. \ref{fig:ModelBalact} the total costs of activation $C_{\text{saldo}}(t)$ can be calculated. Using (\ref{eq:AEP}) the price $P_{\text{bal}}$ can then be calculated based on $C_{\text{saldo}}(t)$ and $RZ_{\text{saldo}}(t)$. Fig. \ref{fig:Modelledbalancingcosts} shows the results of the simulation in comparison to the effective data over the last three years. Compared to the intra-day results, where the cost of negative and positive power is significantly different, the cost structure of balancing energy is almost symmetric and rises similar for negative and positive balancing. The cost uncertainty is larger than the uncertainty as a result of a more complex and harder to train model.

%% END OF FILE

\section{results}\label{sec:results}
In the following section, we will discuss the results of our analysis. Profile service costs are additional costs for system operation and thus, for the end consumer, the overall goal must be the reduction of those. Based on our analysis, the following measures may be used to reduce the profile service costs:

\begin{enumerate}
\item{the reduction of the forecast error $\epsilon(t)$}
%\item{the introduction of a negative bias to $\epsilon(t)$}
\item{selection between available markets}
\item{the shut down of intermittent supply}
\end{enumerate}

\subsection{Reduction of the Forecast Error}
\begin{figure}[!htb]
\centering
\includegraphics[width=0.65\textwidth, draft=false]{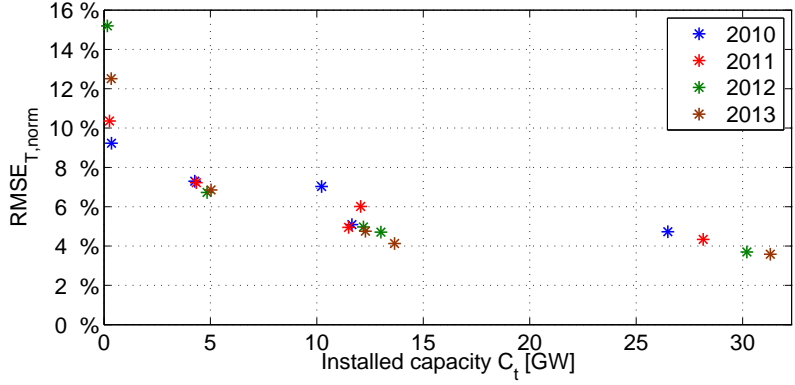}
\caption{\ac{rmse} of $\epsilon(t)$ for wind generated power in Germany normalized by the installed capacity $C_{TSO}$ per \ac{tso}.}
\label{fig:winderror}
\end{figure}

\begin{figure}[!htb]
\centering
\includegraphics[width=0.65\textwidth, draft=false]{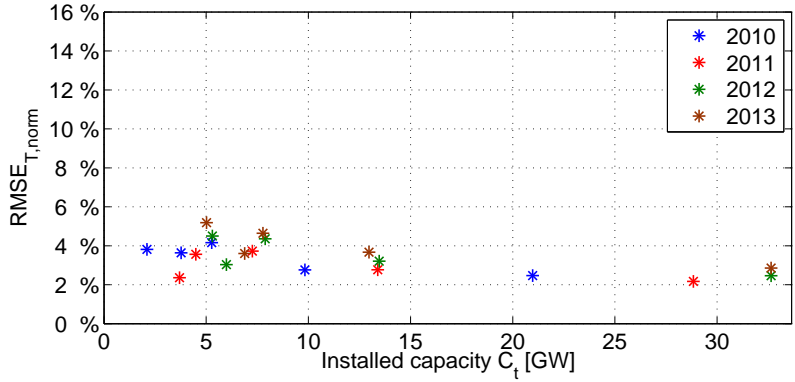}
\caption{\ac{rmse} of $\epsilon(t)$ for \ac{pv} generated power in Germany normalized by the installed capacity $C_{TSO}$.}
\label{fig:pverror}
\end{figure}

The forecast error $\epsilon(t)$ depends mainly on three parameters: First, the prediction horizon, second, the quality of the prediction model and third, the spatial distribution of intermittent supply. Because of the market environment in Germany, the prediction horizon is fixed to 16h to 40h \cite{Lenzi2013}. To prove the influence of the parameters on $\epsilon(t)$, the \ac{rmse} of $\epsilon(t)$ for the years~2010 till~2013 has been normalized by the generation capacity $C_TSO$ to $\text{RMSE}_{TSO,\text{norm}}$ and is given by
\begin{equation}
\text{RMSE}_{TSO,\text{norm}} = \frac{\sqrt{\frac{1}{n}\sum_{t=1}^n\epsilon(t))}}{C_{TSO}},
\end{equation}
where $C_{TSO}$ is the average installed wind \& \ac{pv} generation capacity in the \ac{tso} area for the given year \cite{WindDE},\cite{KraftwerkeDE}. The resulting $\text{RMSE}_{TSO,\text{norm}}$ are shown in Fig. \ref{fig:winderror} and Fig. \ref{fig:pverror}.

The figures confirm that the latter two parameters do have a significant impact on $\epsilon(t)$. First, the introduction of more accurate prediction models for wind generation reduces the relative error for wind generation over time. The blue stars in Fig. \ref{fig:winderror}, representing the prediction error for 2010 per \ac{tso} are significantly higher than the prediction in later years. This result is not significant for \ac{pv} as shown in Fig. \ref{fig:pverror}. Second, the aggregation of larger installed capacities reduces the relative prediction error $\epsilon(t)$ as shown in Fig. \ref{fig:winderror}. This result is also not significant for \ac{pv}.\\
The two measures decrease relative profile service costs for big market actors or big balancing groups. Expensive forecast models can be covered by many installations, improving the prediction quality in combination with the reduced $\text{RMSE}_{TSO,norm}$ due to the big balance group size. Independent of the support schemes, big single actors are able to dispatch \ac{res} generated power more efficiently than small ones and save costs for the end consumer. This is a result of, first, the general larger number of production facilities and therefore better statistical properties and second local disturbances affect the system less.

%\subsection{Introduce Forecast Bias}
%The Analysis in Section \ref{sec:analysis} shows, that the intra-day market costs are non-symmetrical. Market actors are therefore able to reduce balancing costs by conscious negative bias settings of $\epsilon(t)$. This effect is 

\subsection{Selection Between Available Markets}
For big balancing areas like the \ac{res} balancing area from the German \acp{tso}, it could be shown that there is a correlation between trading volumes and costs. Therefore portions of $V_{\text{intraday}}(t)$ and $V_{\text{bal}}(t)$ in (\ref{eq:volumes}) can be selected to reduce $C(t)$ in (\ref{eq:costs}).

\begin{figure}[!htb]
\centering
\includegraphics[width=0.65\textwidth, draft=flase]{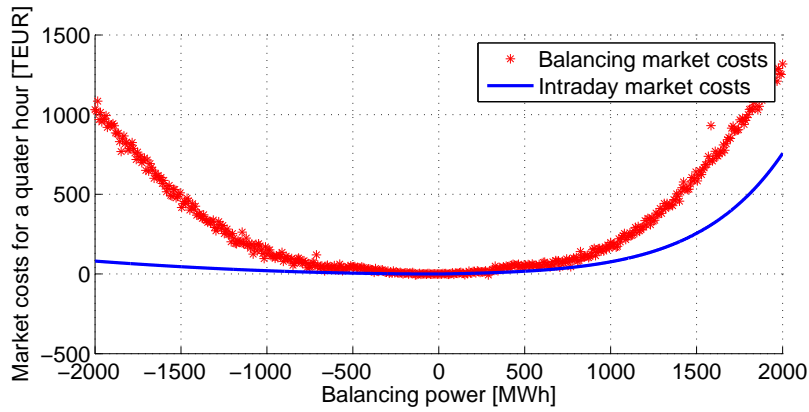}
\caption{Simulated intra-day market costs versus simulated costs of balancing energy.}
\label{fig:modintramodbal}
\end{figure}

\begin{figure}[!htb]
\centering
\includegraphics[width=0.65\textwidth, draft=false]{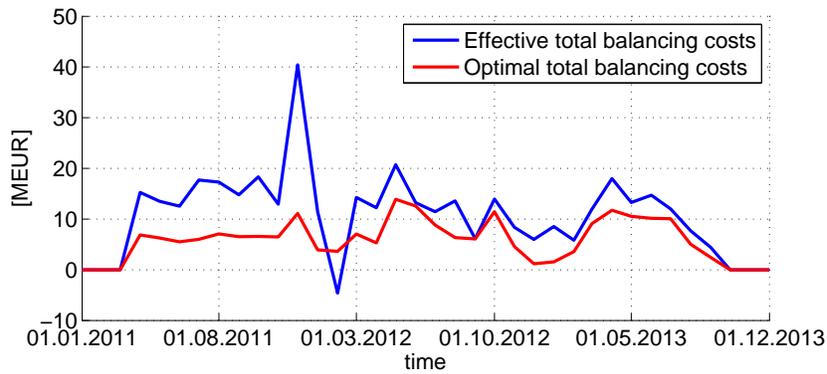}
\caption{Monthly optimized profile service costs for \ac{tso} balancing area compared to effective results.}
\label{fig:optimizer}
\end{figure}

Fig. \ref{fig:modintramodbal} shows the modelled price curves for a 15 minute product on the intra-day market in comparison to the expected costs for balancing energy. Fig. \ref{fig:optimizer} shows the theoretical optimum of balancing cost versus the achieved optimum. The theoretical optimum is significantly lower than the \ac{tso} procurement of profile service costs for the \ac{res} balancing.

The results show, that for market actors it is possible to optimize themselves against the market, even though it is prohibited by law for balancing areas to stay long or short by choice before the activation of balancing reserve.

\subsection{Shut down of intermittent supply}
\begin{figure}[!htb]
\centering
\includegraphics[width=0.65\textwidth, draft=false]{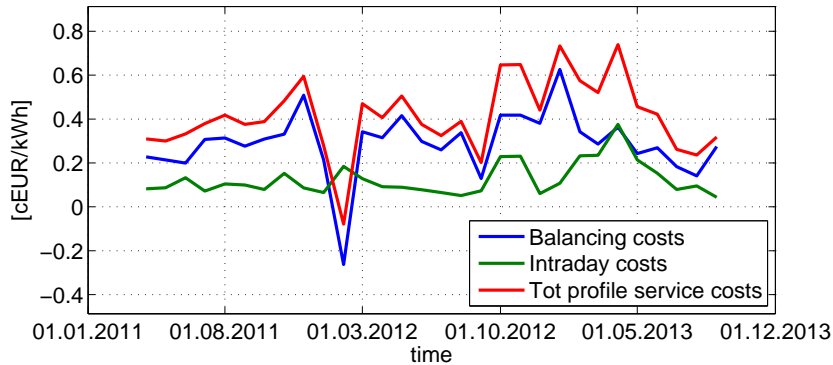}
\caption{Normalized profile service costs for the German \acp{tso}.}
\label{fig:Marginal}
\end{figure}

Profile service costs of the \ac{res} balancing area can be normalized by the dispatched energy given by
\begin{equation}
C_{\text{norm}}(t) = \frac{C(t)}{W_{\text{TSO}}(t)+R_{\text{TSO}}(t)},
\end{equation}
where $C_{\text{norm}}(t)$ is the marginal cost of market actors per sold kWh. If marginal costs of generation are higher than the expected revenues from the market, generation capacity has to be shut down to avoid short term losses for the operator. Because of the support schemes structure in Germany, \ac{res} installations still generate profits due to premium payments, even for hours when marginal costs are higher than the \ac{epex} day-ahead market prices. Fig. \ref{fig:Marginal} shows the effective monthly average profile cost for the \acp{tso}. The sale of \ac{res} generation on the \ac{epex} day-ahead market under its marginal costs is therefore for certain hours, where the \ac{epex} day-ahead market price is lower than $C_{\text{norm}}(t)$, economically inefficient.\\

\begin{figure}[!htb]
\centering
\includegraphics[width=0.60\textwidth, draft=false]{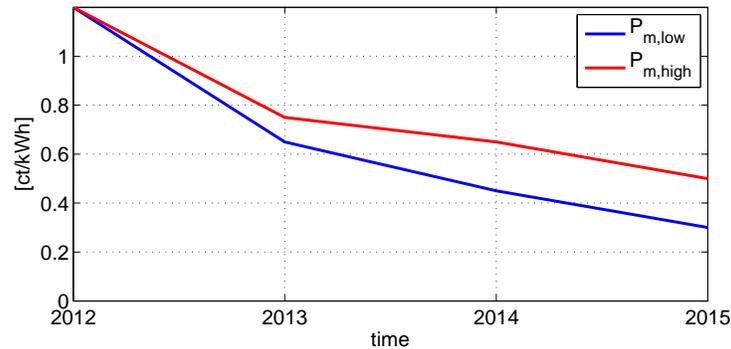}
\caption{Management premium payments for direct marketing of \ac{res}~\cite{Vorschriften2012}.}
\label{fig:Pm}
\end{figure}

\ac{res} installation owners are free to choose every month between direct marketing or fixed \ac{fit}. Market actors using direct sale for their \ac{res} generated power have to pay a premium compared to \ac{fit} to get the assignment. Fig. \ref{fig:Pm} shows the development of management premium payments $P_m$ for wind \& \ac{pv} installations for 2012 to 2015. $P_{m,\text{high}}$ is paid to controllable installations and $P_{m,\text{low}}$ is paid to non-controllable installations. $P_m$ payments are decreasing for the future whereas the realized profile service costs of the \ac{tso} \ac{res} balance group are already rising for certain months above the $P_m$ payments. It therefore can be expected, that operators profile service costs rise above the $P_m$ payments and wind installation owner move back to the \ac{fit} support scheme.

%% END OF FILE
\section{Conclusion}
This paper presented an analysis of the profile service cost as additional marginal costs for \ac{res} deployment. We presented a model to approximate the profile service costs for the \acp{tso} and provide recommendations how to lower the profile service cost via better predictions especially in the case of wind production.

\bibliographystyle{IEEEtran}
% \bibliography{bib/endNoteLibrary_MH}
% \bibliography{bib/MH_Mend_Bib}
\bibliography{MH_Mend_Bib}

\end{document}